\documentclass[%
 aip,
 amsmath,amssymb,
 reprint,%
]{revtex4-1}

\usepackage{graphicx}
\usepackage{dcolumn}
\usepackage{bm}

\usepackage[mathlines]{lineno}
\relax 

\usepackage[utf8]{inputenc}
\usepackage[T1]{fontenc}
\usepackage{mathptmx}
\usepackage{lineno}    
\usepackage{calrsfs}
\usepackage{multirow}  
\usepackage{booktabs}  
\DeclareMathAlphabet{\pazocal}{OMS}{zplm}{m}{n}
\usepackage{xcolor}
\usepackage{relsize}

\usepackage[normalem]{ulem}
\usepackage{amsmath}
\usepackage{amssymb}

\definecolor{LightCyan}{rgb}{0.88,1,1}
\definecolor{piggypink}{rgb}{0.99, 0.87, 0.9}
\definecolor{applegreen}{rgb}{0.55, 0.71, 0.0}
\definecolor{darkpastelgreen}{rgb}{0.01, 0.75, 0.24}
\definecolor{green-yellow}{rgb}{0.68, 1.0, 0.18}

\newcommand{\beq}{\begin{equation}}
\newcommand{\eeq}{\end{equation}}
\newcommand{\beqa}{\begin{eqnarray}}
\newcommand{\eeqa}{\end{eqnarray}}

\begin{document}
\preprint{AIP/123-QED}

\title[Absolute measurement of the Fano factor using a Skipper-CCD]{Absolute measurement of the Fano factor using a Skipper-CCD}

\author{Dario Rodrigues}
\affiliation{\normalsize\it 
Department of Physics, FCEN, University of Buenos Aires and IFIBA, CONICET, Buenos Aires, Argentina}
\affiliation{\normalsize\it 
Fermi National Accelerator Laboratory, PO Box 500, Batavia IL, 60510}

\author{Kevin Andersson}
\affiliation{\normalsize\it 
Department of Physics, FCEN, University of Buenos Aires and IFIBA, CONICET, Buenos Aires, Argentina}
\affiliation{\normalsize\it 
Fermi National Accelerator Laboratory, PO Box 500, Batavia IL, 60510}

\author{Mariano Cababie}
\affiliation{\normalsize\it 
Department of Physics, FCEN, University of Buenos Aires and IFIBA, CONICET, Buenos Aires, Argentina}
\affiliation{\normalsize\it 
Fermi National Accelerator Laboratory, PO Box 500, Batavia IL, 60510}

\author{Andre Donadon}
\affiliation{\normalsize\it 
Department of Physics, FCEN, University of Buenos Aires and IFIBA, CONICET, Buenos Aires, Argentina}
\affiliation{\normalsize\it 
Fermi National Accelerator Laboratory, PO Box 500, Batavia IL, 60510}

\author{Ana Botti}
\affiliation{\normalsize\it 
Department of Physics, FCEN, University of Buenos Aires and IFIBA, CONICET, Buenos Aires, Argentina}

\author{Gustavo Cancelo}
\affiliation{\normalsize\it 
Fermi National Accelerator Laboratory, PO Box 500, Batavia IL, 60510}

\author{Juan Estrada}
\affiliation{\normalsize\it 
Fermi National Accelerator Laboratory, PO Box 500, Batavia IL, 60510}

\author{Guillermo Fernandez-Moroni}
\affiliation{\normalsize\it 
Fermi National Accelerator Laboratory, PO Box 500, Batavia IL, 60510}

\author{Ricardo Piegaia}
\affiliation{\normalsize\it 
Department of Physics, FCEN, University of Buenos Aires and IFIBA, CONICET, Buenos Aires, Argentina}

\author{Matias Senger}
\affiliation{\normalsize\it 
Department of Physics, FCEN, University of Buenos Aires and IFIBA, CONICET, Buenos Aires, Argentina}
\affiliation{\normalsize\it 
Fermi National Accelerator Laboratory, PO Box 500, Batavia IL, 60510}

\author{Miguel Sofo Haro}
\affiliation{\normalsize\it 
Fermi National Accelerator Laboratory, PO Box 500, Batavia IL, 60510}
\affiliation{Centro At\'omico Bariloche, CNEA/CONICET/IB, Bariloche, Argentina}

\author{Leandro Stefanazzi}
\affiliation{\normalsize\it 
Fermi National Accelerator Laboratory, PO Box 500, Batavia IL, 60510} 

\author{Javier Tiffenberg}
\affiliation{\normalsize\it 
Fermi National Accelerator Laboratory, PO Box 500, Batavia IL, 60510}

\author{Sho Uemura}
\affiliation{\normalsize\it 
Raymond and Beverly Sackler School of Physics and Astronomy, \\
 Tel-Aviv University, Tel-Aviv 69978, Israel}

\date{\today}

\hfill{FERMILAB-PUB-20-283-E}

\begin{abstract}
Skipper-CCD can achieve deep sub-electron readout noise making possible the absolute determination of the exact number of ionized electrons in a large range, from 0 to above 1900 electrons.
In this work we present a novel technique that exploits this unique capability to allow self-calibration and the ultimate determination of silicon properties.
We performed an absolute measurement of the variance and the mean number of the charge distribution produced by $^{55}$Fe X-rays, getting a Fano factor absolute measurement in Si at 123K and 5.9 keV.
A value of 0.119 $\pm$ 0.002 was found and the electron-hole pair creation energy was determined to be (3.749 $\pm$ 0.001) eV.
This technology opens the opportunity for direct measurements of the Fano factor at low  energies.

\vspace{1mm}
\end{abstract}

\vspace{-1mm}

\maketitle

\section{\label{sec:intro}Introduction}

\vspace{-3mm}
The ratio between the observed statistical fluctuations in the number of charge carriers and that expected from a pure Poisson statistics, the Fano\cite{Fano1947} factor $F$, has been historically used --along with the electron-hole pair creation energy $\epsilon_{eh}$ 
to characterize the response of different detectors to $X$ radiation\cite{Ryan_1973, Alig1983, LECHNER1996, Scholze1998, LOWE2007}.

A precise determination of $\epsilon_{eh}$ has also implications for Dark Matter Searches\cite{SENSEI_2019} and reactor-Neutrino experiments\cite{Aguilar_Arevalo_2019, vIOLETA} to reconstruct the energy deposited by interacting particles in the detector. 
To measure $F$ at low energies is also key in the sensitivity calculation of low-mass dark matter experiments\cite{Durnford2018, Ramanathan2017}. 
For a review of ionization modeling in silicon at low energy and further discussion of $\epsilon_{eh}$ and $F$ see Ref.~\onlinecite{Ramanathan2020}.

Additionally, both $\epsilon_{eh}$ and $F$ have a significant role in the calibration requirements of Charge Coupled Devices (CCD) imaging spectrometers, such as in astronomy\cite{Fraser_1994}, and have been measured in Si using conventional CCDs\cite{Janesick1988, Owens2002, Janesick_2016, Kotov_2018}.
Since these kind of measurements are affected by sensor calibration accuracy, different approaches were proposed to reduce the contribution from systematic uncertainties in gain determination. 
For instance, Kotov et al.\cite{Kotov_2018} have used an optical technique which takes advantage of the Poisson distribution properties. 

One of the main contributions to systematic uncertainty when using a conventional CCD comes from the low-frequency noise ($\sigma_{RN}$), 
which impose a lower limit to the readout noise of nearly $\sigma_{RN} \approx$ 1.8 e$^-$ rms/pix\cite{Janesick_2016, Bebek_2017}.
As a consequence, the  actual variance of the charge distribution ($\sigma^2$) cannot be measured, and instead, a larger $\sigma_{obs}^2=\sigma^2+\sigma_{RN}^2$ is observed.
Such readout noise essentially makes the direct determination of $F$ impossible for low energies where $\sigma_{RN}^2$ cannot be neglected against $\sigma^2$. 
Furthermore, the capability to reconstruct the total charge produced by each event is also affected by charge collection inefficiencies, 
and the difficulty to determine the actual size of each cluster of pixels produced by each interaction, which introduce an extra systematic uncertainties.

Here we report the first measurement of $\epsilon_{eh}$ and $F$ using a Skipper-CCD, which allows to reach sub-electron readout by measuring the charge in each pixel as many times as desired without destroying it. 
Exploiting this feature, we developed a novel method for absolute  self-calibration by identifying the quantized-charge peaks that correspond to all electron multiplicities between 0 and $\sim$1900 e$^{-}$. 
Based on this, taking advantage of the almost perfect Charge Collection Efficiency (CCE) and the sub-electron readout noise, our results represent the most precise measurement of both $\epsilon_{eh}$ and $F$ in Si.


\section{\label{sec:skipper}Skipper-CCD detector}

The main difference between conventional scientific CCDs and Skipper-CCDs lies in the non-destructive readout system of the latter, which allows to repeatedly measure the charge in each pixel.
For uncorrelated samples, this capability results in the reduction of the readout noise in a factor equal to the square root of the number of samples \cite{Tiffenberg2017}.
This feature enables the precise determination of the number of electrons in each pixel, which means that single photon counting is possible in the low energy region (optical and near-infrared).

\begin{table}
\caption{\label{tab:characteristics} Main characteristics of the CCD detector used in this work.}
\begin{ruledtabular}
\begin{tabular}{lcc}
Characteristics             & Value      & Unit      \\
\hline
Format                      & 4126 x 886 & pixels    \\
Pixel size                  & 15         & um        \\
Thickness                   & 200        & um        \\
Operating Temperature       & 123        & K         \\
Readout noise (1 sample)    & 3.5        & e$^-$ rms/pix \\
Readout noise (300 samples) & 0.20       & e$^-$ rms/pix \\
\end{tabular}
\end{ruledtabular}
\end{table}

\begin{figure}
\includegraphics[width=3.3in]{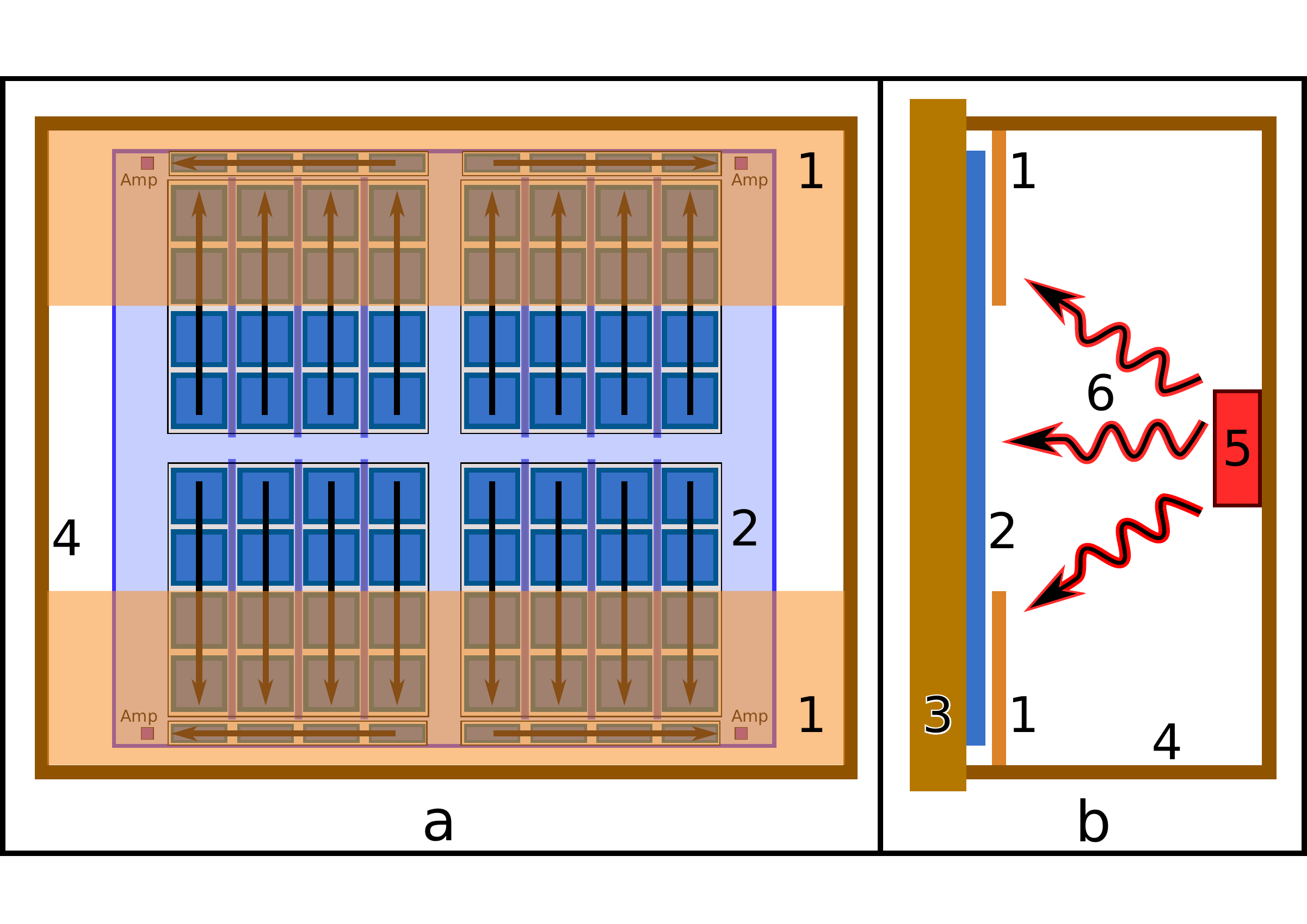}
\caption{\label{fig:setup} Experimental setup. a) Front view. b) Lateral view. Arrows indicate how the charges are moved in each quadrant during readout. 1: Thin Cu foils cover half of the detector near amplifiers to shield them from direct X-ray illumination. 2: Skipper-CCD, 3: Cold Cu piece, 4: Cold Cu box, 5: $^{55}$Fe radioactive source,  6: X-rays.}
\end{figure}

The Skipper-CCD sensor used in this work is a back-illuminated fully-depleted CCD designed by the Microsystems Laboratory at LBNL and fabricated at Teledyne DALSA Semiconductor. Table \ref{tab:characteristics} describes its main characteristics. 
The detector is divided in four quadrants, each of them constituted by 443 rows and 2063 column.  The general structure of the detector pixel array is schematized in Fig.~\ref{fig:setup}a. 

By contrast to others scientific CCD with several microns of dead layer in the back, the Skipper-CCD used in this work has a special back side treatment for photon collection\cite{Holland_2003}.  That side is covered by three thin layers: $\sim$20 nm Indium Tin Oxide (ITO), $\sim$38 nm ZrO$_2$, and $\sim$100 nm SiO$_2$.
The detector was installed in a 30 cm $\times$ 30 cm $\times$ 30 cm aluminum Dewar and it was cooled at 123K using a cryocooler. 
The readout and control systems are fully integrated in a new single-board electronics optimized for Skipper-CCD sensors. This Low-Threshold-Acquisition system (LTA) was developed in-house\cite{Moroni2019} and provides a flexible and scalable solution for detectors with target masses up to a few hundred grams.

\section{Calibration and linearity}
\begin{figure}
\includegraphics[width=3.5in]{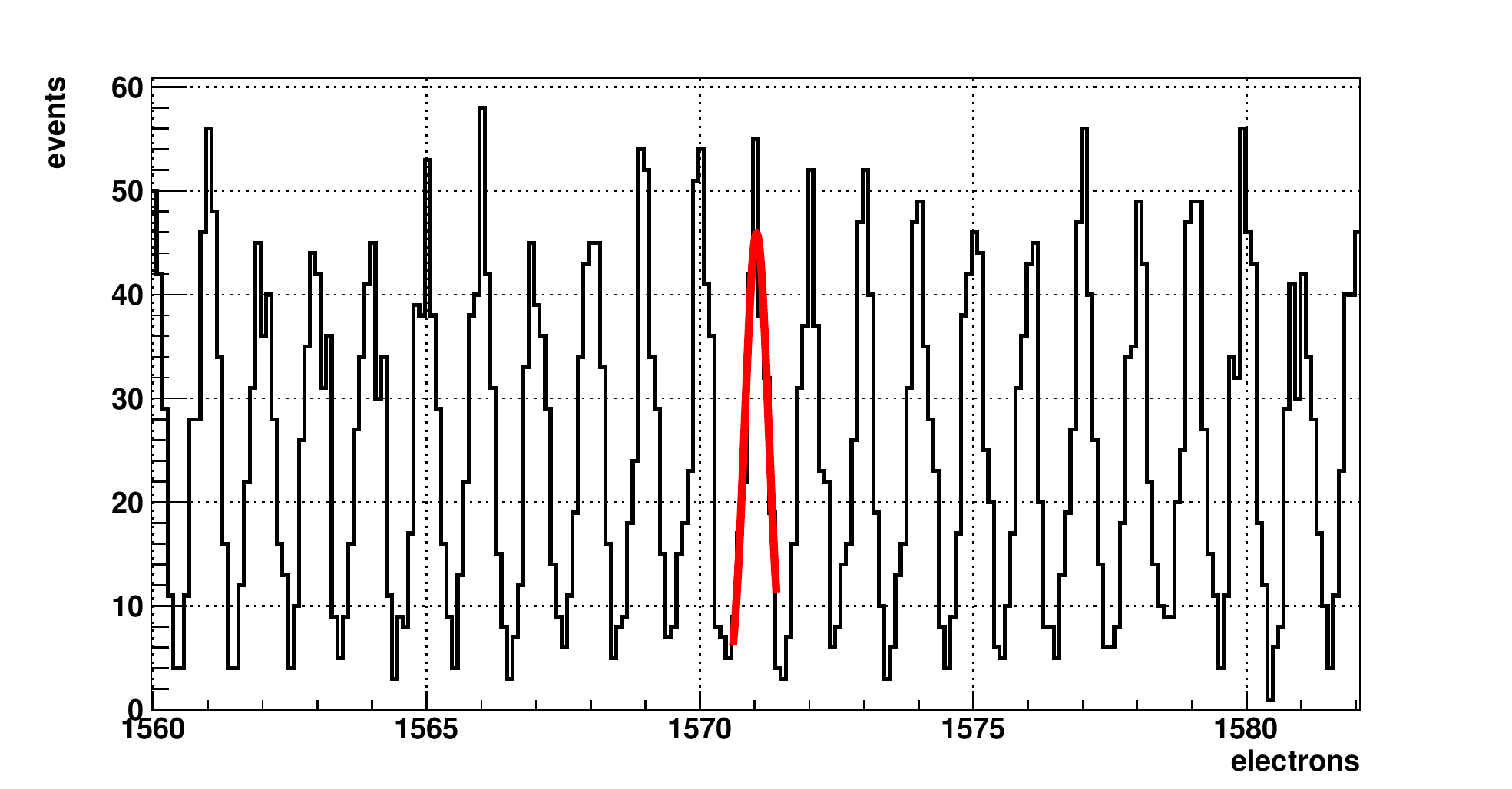}%
\caption{\label{fig:peaks} Self-calibration performed taking 300 samples per pixel. The ADU mean numbers per peak where fitted against  the peak number using a fourth-degree polynomial to take into account for nonlinearities. Here we display a zoom of the full spectrum centered at the electron mean number produced by 5.9 keV X-rays peaks from $^{55}$Fe. All the peaks from 0 to $\sim$1900~e$^{-}$ are clearly identified in the full spectrum.}
\end{figure}

A self-calibration procedure was performed to determine the relationship between the number of electrons in each pixel and the signal readout value in Analog to Digital Units (ADU).
An LED installed inside the Dewar was used to populate the CCD pixels with the electrons produced by 405 nm photons. 
In order to cover a large range of electrons per pixel, we performed several measurements increasing the light exposure time. Thus, we produced different overlapping Poissonian distributions with increasing mean number of electrons.
All these measurements were performed taking 300 samples per pixel. As a result, the readout noise was reduced by a factor $\sqrt{300}\sim$ 17.3, achieving a final value of 0.2 e$^-$. 
This allowed us to distinguish between consecutive peaks in the full range from zero up to 1900 e$^-$. 
The mean value in ADU for each of those peaks was determined by means of Gaussian fits. Then, we completed the self-calibration by simple assignment of each ADU mean value to the corresponding peak number, {\it i.e.} the number of electrons.
Fig.~\ref{fig:peaks} presents the peaks in the region between 1560 and 1582 electrons.

\begin{figure}[ht]
\includegraphics[width=3.5in]{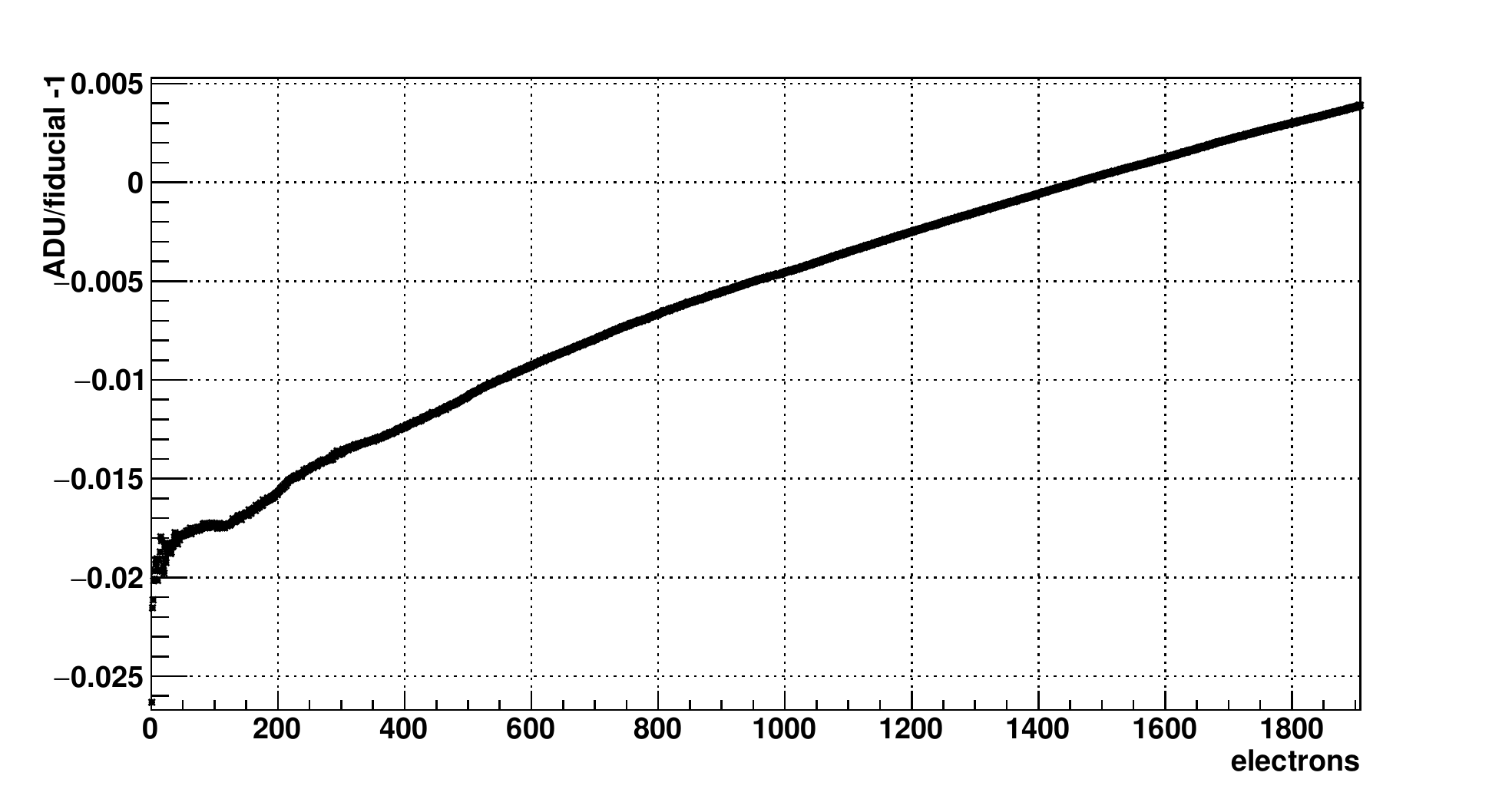}
\caption{\label{fig:linearity} Nonlinearities in one of the four Skipper-CCD readout electronics. The $y$ axis plots the fractional change in the number of electrons from calibration relative to the actual one, perfectly linear readout electronics would yield zeros.}
\end{figure}

In regards to the nonlinearities in the readout electronics, Fig.~\ref{fig:linearity} displays departures from 1 of the ratio between the number of electrons calculated from a linear self-calibration and the actual number of electrons per pixel.
In contrast with the usual nonlinearity measurements in  conventional CCD (see, for instance, Fig. 1 in  Ref.~\onlinecite{Bernstein_2017}), Skipper-CCDs allow to quantify nonlinearities for all occupancies in a full range.

\section{Measurements}

The X$_K$ rays emitted after the $^{55}$Fe electron capture decays are widely used for CCD calibration \cite{Janesick_2016}.
Their energies, known with excellent precision, are summarized in Table~\ref{tab:table1}.
For our purpose, we used an electroplated $^{55}$Fe radioactive source with a diameter of $\sim$5 mm and an activity of $\sim$0.1 $\mu$Ci.
This source was placed facing the backside of the CCD and $\sim$40 mm apart as depicted in Fig.~\ref{fig:setup}b.

\begin{table}
\caption{\label{tab:table1} $^{55}$Fe X-rays energies and intensities\cite{Krause1979, TabRad_v3} of interest in this work. Source: https://xdb.lbl.gov/}
\begin{ruledtabular}
\begin{tabular}{ccc}
X$_K$ & Energy (keV)  & Rel Intensity       \\
\hline
$\alpha_2$      & 5887.6 & 8.5  (4)     \\ 
$\alpha_1$      & 5898.8 & 16.9 (8)    \\
$\beta_{3}$     & 6490.4 & 3.4  (11)     \\
\end{tabular}
\end{ruledtabular}
\end{table}

In a significant fraction of the $^{55}$Fe decays, the energy is transferred to an orbital electron instead of an X-ray. These Auger electrons leave the atom with an energy just a few eV lower than the X-rays due to the ionization energy, and create a continuous energy spectrum when they hit the CCD. 
To avoid this background we covered the $^{55}$Fe source with a 20 $\mu$m Mylar foil that stops the keV-electrons and has very small probability of producing small-angle Compton scattering of the X-rays. 
The total thickness of dead layers ($\sim$160 nm), however, introduces a probability of interaction for 5.9 keV photons of $\sim$1.4\%.

\paragraph{Data acquisition procedure}\label{par:data}

To reduce impact of dark current we limit the exposure/readout time by simultaneously reading the 4 quadrants of the CCD and restricting the acquisition to only 50 rows per quadrant. Each row containing 500 pixels (7 prescan, 443 active, and 50 overscan). Thus, we took images with an active area of 22150 pixels. The exposure to the X-rays was done moving quickly the charges in those pixels ($\sim$30 seconds of effective exposure) in order to get a X-rays hitting rate of $\sim$4 Hz in the active area of each image, turning in $\sim$120 events on average.
We performed 300 samples per pixel, which corresponds to a readout time of $\sim$10 minutes per image. After readout, the 300 samples taken for each pixel are averaged and the empty pixels in the overscan are used to compute and substract a baseline for each row. The resulting image contains 443x50 pixels for each quadrant and the measured charge is represented in ADUs that is converted into electrons using the self-calibration procedure described above.

At the end of each exposure/readout cycle, all the charge collected by the CCD during this time is flushed in a quick clean procedure that takes about a second.

Because of the relatively high rate of X-rays photons hitting the CCD, we covered half of each quadrant with a thin Cu foil in the region close to the amplifiers (see Fig.~\ref{fig:setup}). 
In this way, we have exposed the uncovered area to the X-rays while the charge is quickly moved under the Cu foils where, shielded from the source, they wait to be read.

\paragraph{Black body radiation shield}

To minimize backgrounds produced by infra-red (IR) photons emitted by the inner surfaces of the vacuum vessel, which is at room temperature, we covered the detector with a cold copper box as shown in Fig.~\ref{fig:setup}b. 
This box is in thermal contact with the cold copper piece in which the detector is mounted and shields it from black body radiation originating on the surrounding walls.

\section{DATA ANALYSIS AND RESULTS}

\begin{figure}
\includegraphics[width=3.4in]{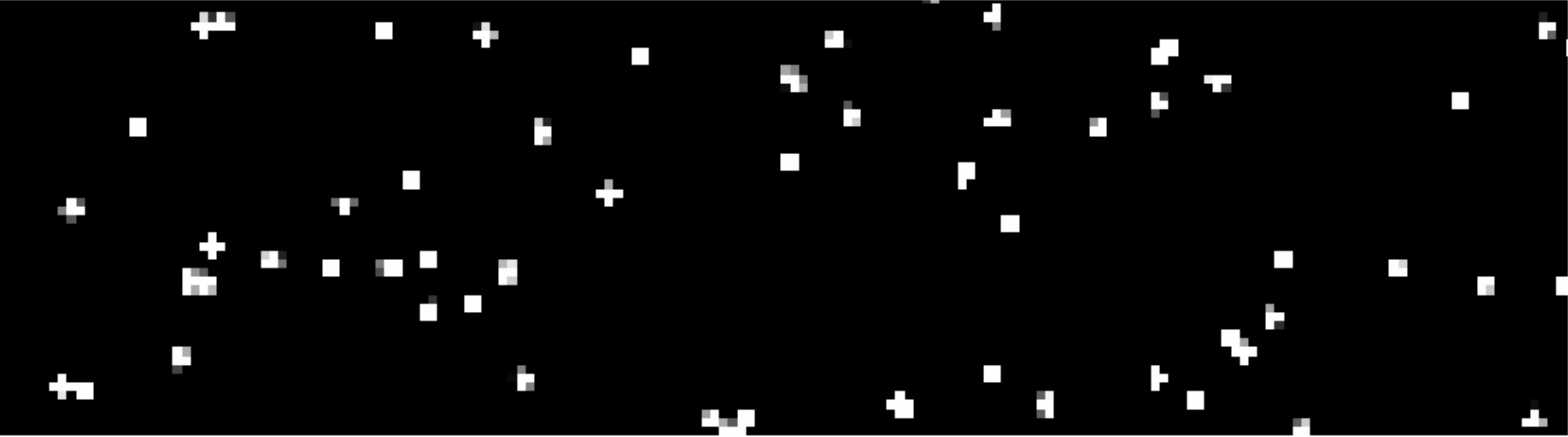}
\caption{\label{fig:imagen} Section of a Skipper-CCD image after exposing it to the X-rays emitted by the $^{55}$Fe radioactive source where different cluster size can be identified.}
\end{figure}

\paragraph{Event reconstruction}
The images recorded using the procedure described above contain ionization events produced by X-rays from the $^{55}$Fe source and other environmental radiation. 
A fraction of a typical image is presented in Fig.~\ref{fig:imagen}. 
As the Skipper-CCD used for this work was back-illuminated by the X-rays from the source, the resulting interactions mostly took place in the first 30~$\mu$m of the back-side of the CCD. 
Due to charge diffusion during the charge collection process, the charge of the resulting events is distributed among several pixels following a 2D Gaussian distribution~\cite{Haro2020}.
The total number of electrons generated by each X-ray event was reconstructed by running a clustering algorithm in which all non-empty neighboring pixels are group together and are considered to be part of a the same event. 

It is worth noting that Skipper-CCD enables the charge determination in clusters of different sizes, introducing a pretty small systematic uncertainty in the estimation of the total number of electrons produced by each event.
With a readout noise of 0.2 e$^-$, the boundaries of each cluster can be determined with a probability of miss classification as low as $p$ = 0.062 per external surrounding pixel. 
In addition, this also implies that this measurement is robust to charge transfer inefficiencies that may spread the charge among neighboring pixels. 
The probability of any electron from an event being separated from the other electrons by one or more empty pixels is essentially null for all practical purposes.

\paragraph{Quality cuts}

To reject merged events, clusters with relatively large or small variance in any of both x and y directions were discarded and only relatively circular clusters, compatible with the expected 2D Gaussian shape~\cite{Haro2020}, are selected.
To reject Compton events produced in the bulk of the CCD by high-energy environmental radiation we place a cut on the size of the clusters to select events produced by interactions in the first 30~$\mu$m of the backside of the CCD.
The clusters size distribution after these cuts has a mean value of (12.4 $\pm$ 2.7 ) pixels.

\paragraph{Readout noise}\label{par:readout_noise}
The probabilities of counting one more and one less electron (with respect to the actual number) in the pixels building clusters are equal to $p$. 
Taking into account the cluster size distribution, eventual inner or external miss classification introduces a bias as low as 0.1 e$^-$ and a readout noise $\sigma_{RN}$ = 0.5 e$^-$ rms/cluster. 
Nevertheless, $\sigma_{RN}$ constitutes the main contribution to the uncertainty of $\epsilon_{eh}$.

\paragraph{Dark Current Effect}

The Dark Current (DC) was measured at the same experimental conditions but without the $^{55}$Fe radioactive source. It was computed as the ratio between 1-electron events and empty pixels, resulting in $\sim$1$\times 10^{-5}$ electron per pixel per second. 
Thus, taking into account the mean size of clusters, we expect only (0.04 $\pm$ 0.01) extra electrons due to the DC during the 10 minutes spent in exposing and reading each image. Therefore, the DC effect can be neglected without introducing a significant bias in our results.

\paragraph{Charge Collection Efficiency}

There are two effects responsible for degrading CCE: recombination and charge transfer inefficiencies. As already discussed, the latter is insignificant when using Skipper-CCD.
The cryogenic temperature of operation, the very low-doped silicon and the high electric field ($\sim$350 V/mm) in the bulk of the CCD prevent signal charge to be lost by recombination. 
As a result, in the active volume of a fully depleted CCD detector, CCE can be considered essentially as one for all practical purposes.
Back-illuminated CCDs in astronomy (such as the one used in this work) are treated to have a thin entrance window for light, with low reflectivity \cite{JPLdelta_1994,JPLdelta_2016,DESICCD2017}.
Ref.~\onlinecite{LBNLQE} compares the detection efficiency for visible photons with the reflectivity of the backside of  a CCD sensor that was produced using the same fabrication process (and the identical design for the active region) as the one used in this work.
These studies show that all photons with wavelengths (absorption lengths) between 650~nm and 850~nm are fully detected unless they are reflected on the back surface. 
This is only possible if the bulk of the detector has a CCE close to 100\%. There may be a small spectral distortion due to partial charge collection near the backside of the CCD~\cite{Fernandez-Moroni:2020abn}, but this effect only affect events interaction in the first micron and has a negligible impact on the shape of the peak as it shows a continuous almost flat spectrum (between 0 eV and the energy of the peak) with an amplitude that is almost four orders of magnitude smaller than the K$_{\alpha}$ peak.

\paragraph{Traps}

CCD could suffer the presence of traps able to introduce irregularities during the readout process. 
Since traps release charges with exponential distribution in time, those columns are expected to have a higher number of one-electron events (\textit{hot columns}). 
To ensure our results is not biased by this process, we have identified and masked all the hot columns out. However, following this very conservative rule did not significantly impact in our final results.

Furthermore, as a sanity check, we have also computed the Fano factor for 20 intervals of 20 columns each. As a result, we have observed the expected fluctuation about the global value without outliers. This also proves that there is not a significant effect because of traps along the serial register.
Besides, we have compared the number of charges per pixel in the first and second measurement in the sense node ensuring that no charges are missing during the skippering steps.

\paragraph{Unbinned multipeaks fit}

K$_{\alpha}$ and K$_{\beta}$ X-rays peaks were fitted using the likelihood given by Eq.~(\ref{eq:likelihood}). It is the result of the convolution of two exponential with one Gaussian distribution for each of the three peaks given in Table \ref{tab:table1} (for a detailed derivation see Ref.~\onlinecite{BORTELS1987}). Here, this is just an empirical fit that is not motivated by the physical reason found in alpha-particle spectroscopy.

\begin{widetext}
\begin{eqnarray}
\label{eq:likelihood}
{\cal L}(e|\mu_1,\mu_3, \sigma_1,\lambda_1,\lambda_2, \eta_1=\eta_2, \eta_3)= 
\mathlarger{\sum}_{j=1}^3 I_j \Biggl[
\eta_j \frac{\lambda_1}{2} && \exp \Bigl[(e-\mu_j)\lambda_1 + \frac{\sigma_j^2 \lambda_1^2}{2} \Bigr]\times 
\text{Erfc}\Bigl[\frac{1}{\sqrt2} \Bigl(\frac{e-\mu_j}{\sigma_j}+\sigma_j \lambda_1 \Bigr)\Bigr] + \nonumber\\
(1-\eta_j) \frac{\lambda_2}{2} && \exp \Bigl[(e-\mu_j)\lambda_2 + \frac{\sigma_j^2 \lambda_2^2}{2} \Bigr]\times
\text{Erfc}\Bigl[\frac{1}{\sqrt2} \Bigl(\frac{e-\mu_j}{\sigma_j}+\sigma_j \lambda_2 \Bigr)\Bigr]\Biggr]
\end{eqnarray}
\end{widetext}

\noindent where $\mu_j$, $\sigma_j$ and $I_j$ represent the mean number, the standard deviation, and the relative intensity of each peak $j$ with energy $E_j$.
$\lambda_1$ and $\lambda_2$ stand for the parameters of the two exponential distributions convoluted with each Gaussian, while $\eta_j$ sets the relative weight between those exponential.

Since the difference between the energy of K$_\alpha$ peaks is only 11.1 eV (see Table \ref{tab:table1}), we can safely assume the same $\epsilon_{eh}$ and $F$ for both of them. Therefore, we set $\mu_2=\mu_1 \times E_2/E_1$ and  $\sigma_2=\sigma_1 \times \sqrt{E_2/E_1}$.
In the case of K$_\beta$ peak, we also assume the same $F$ but we allowed for $\epsilon_{eh}$ to take a different value. These conditions are fulfilled by letting $\mu_3$ be a free parameter (see Eq. (\ref{eq:likelihood})), and fixing $\sigma_3=\sigma_1 \times \sqrt{\mu_3/\mu_1}$.

\begin{figure}
\includegraphics[width=3.7in]{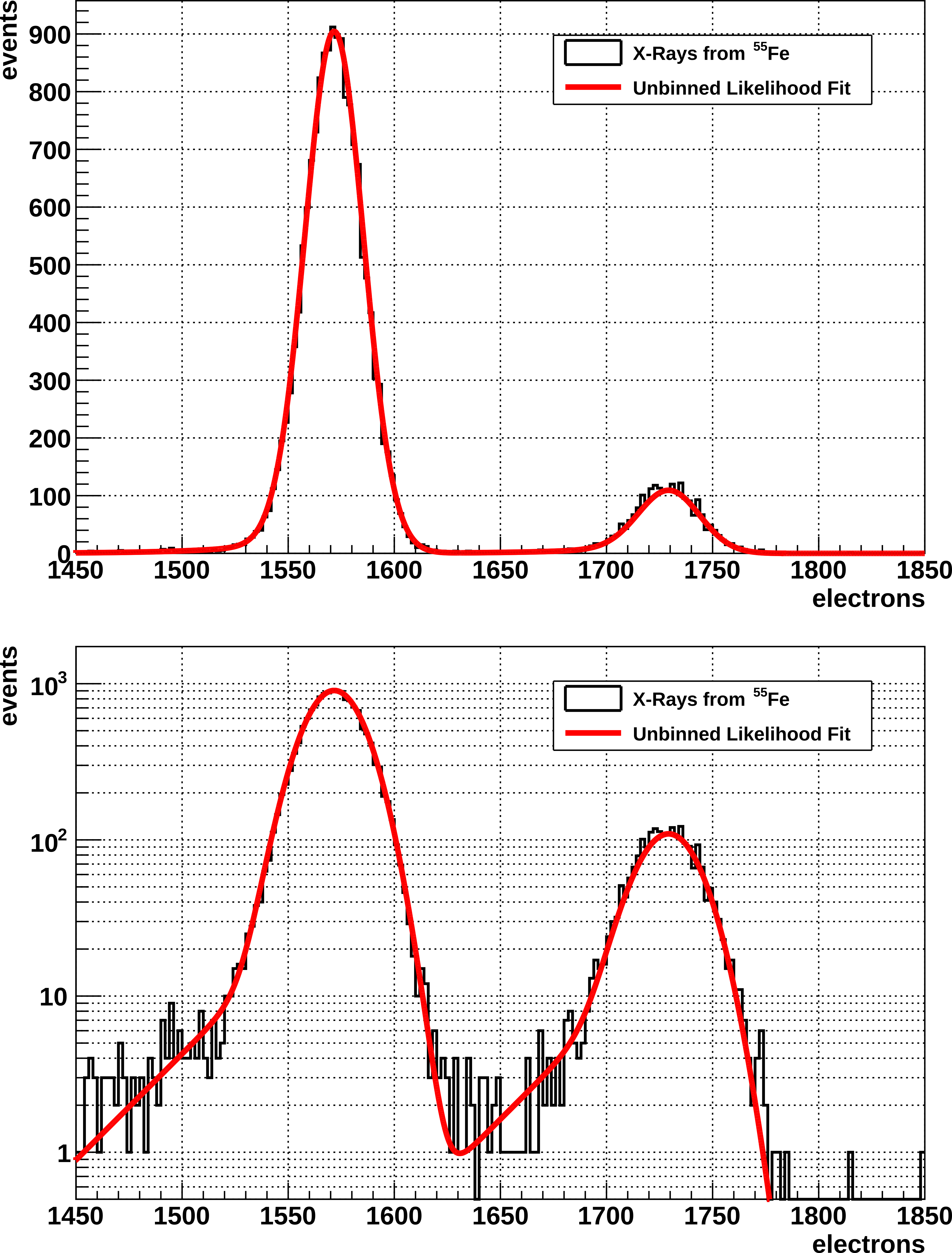}
\caption{\label{fig:peak_at_5p9} X-ray peaks at 5.9 keV and 6.5 keV. Red line corresponds to the Unbinned multipeaks likelihood fit.}
\end{figure}

Fig.~\ref{fig:peak_at_5p9} presents the unbinned likelihood fit for the X-rays peaks over a total of 18085 events after selection and quality cuts. The relevant fitted parameters and the values for $\epsilon_{eh}$ and $F$ are listed in Table \ref{tab:fitted_parameters}.

The fitted values for $\mu_1$, $\mu_3$ and $\sigma_1$ are very robust against changes in the energy range considered for fitting and therefore, so are the computed for $\epsilon_{eh}$ and $F$. 
Such  changes only affect $\lambda_1$, $\lambda_2$, and $\eta_j$ values, which essentially fulfill the function of accounting for the small left tails in Fig.~\ref{fig:peak_at_5p9}. 

\begin{table*}
\caption{\label{tab:fitted_parameters} Fitted parameters $\mu$ and $\sigma$ obtained by maximizing the Likelihood given by Eq.(\ref{eq:likelihood}). $\Delta\mu$ and $\Delta\sigma$ are their corresponding fitting uncertainties. $\Delta$F is the absolute uncertainty in $F$ after considering both systematic and statistics contribution.  $\epsilon_{eh}$ and its uncertainty are also presented.
Fano factor was set to be the same for all peaks.
$\epsilon_{eh}$ was set to be the same for $\alpha$ peaks.
The result fitting only the $\beta$ peak is also included.}
\begin{ruledtabular}
\begin{tabular}{ccccccccc}
 X$_{K}$                & $\mu$   & $\Delta \mu$    & $\sigma$ & $\Delta \sigma$ & $F$ & $\Delta F$  & $\epsilon_{eh}$ & $\Delta \epsilon_{eh}$ \\ 
 \hline
 $\alpha_2$    & 1570.50 & 0.18 & 13.68 & 0.12 &  \multirow{3}{*}{0.119} & \multirow{3}{*}{0.002}&\multirow{2}{*}{3.749} & \multirow{2}{*}{0.001}  \\
 $\alpha_1$    & 1573.48 & 0.18 & 13.69 & 0.12 &                          &                                              &                         \\
 $\beta_{3}$   & 1730.50 & 0.55 & 14.36 & 0.13 &                          &                       & 3.751                &                   0.002  \\
\end{tabular}
\end{ruledtabular}
\end{table*}

\paragraph{Systematic uncertainties}
As mentioned above, due to miss classification a systematic error is originated in the cluster readout noise ($\sigma_{RN}$=0.5 e$^-$). This was summed in quadrature with $\sigma$ turning in an insignificant contribution. Additionally, there is a contribution of 0.010 on the Fano factor because of the quality cuts applied on the energy of events to restrict the fitting domain, and 0.005 originated in the selection of minimum and maximum variance for a cluster to be accepted. The last two were summed in quadrature with the uncertainty obtained from propagation of $\Delta\sigma$ and $ \Delta\mu$, the fitting uncertainties for $\sigma$ and $\mu$ respectively (see Table \ref{tab:fitted_parameters}) through the Fano factor formula. 
As a result, a total absolute uncertainty of $\Delta$F = 0.0023 was determined.


\section{DISCUSSION AND CONCLUSIONS}

Our results are in agreement with two pioneer works: Ryan et al.\cite{Ryan_1973}, who almost fifty years ago have reported $\epsilon_{eh}$ = 3.745 $\pm$ 0.003 eV, and Alig et. al.\cite{Alig1983}, who almost forty years ago using Monte Carlo simulations found $F$ = 0.113 $\pm$ 0.005. 
However, going to more recent works, we have found large discrepancies with other published results\cite{Janesick1988, Fraser_1994, Owens2002}, where $F$ values between 0.14 and 0.16 have been informed.

Although initially $\epsilon_{eh}$ and $F$ were treated as material constants\cite{Ryan_1973, Alig1983}, since then, many authors have investigated experimentally\cite{PEHL1968, LECHNER1996, Scholze1998, LOWE2007} and by means of Monte Carlo simulations\cite{Fraser_1994, MCCARTHY1995, MAZZIOTTA2008} their dependence on both energy and temperature. 
As a result, nowadays we know that, $\epsilon_{eh}$ decreases as the temperature or the energy increases, while $F$ changes in a lesser extent.

Kotov et al.\cite{Kotov_2018}, using conventional CCD at 185K have reported $\epsilon_{eh}$=(3.650 $\pm$ 0.009) eV and $F$=0.128 $\pm$ 0.001. According to the gradient informed by Lowe et al.\cite{LOWE2007}, the disagreement with our results can only be partially explained as a consequence of the difference in temperature.
We do observe a perfect agreement with the values published by Lowe et al.\cite{LOWE2007}. They have measured $\epsilon_{eh}$ as a function of temperature, and according to what they have reported, $\epsilon_{eh}$ =(3.743 $\pm$ 0.090) eV and $F$ = 0.118 $\pm$ 0.004 is what we should expect at 123K.

We've shown how the sub-electron readout noise achieved by Skipper-CCDs enables its self-calibration.
It allowed us to perform an absolute determination of the variance and mean number of the charge distribution produced by X-rays from $^{55}$Fe.
Thus, we've obtained the first Fano factor absolute measurement in Si and the most precise determination of both the electron-hole pair creation energy $\epsilon_{eh}$ and $F$.

A notable feature of our results is that thanks to the Skipper-CCD performance they were possible with neither subtracting reading noise nor correcting for CCE, as it was needed in previous work.

Ongoing experiments using X-rays from Al and F fluorescence will exploit the capability of this technology for probing the theoretically expected Fano factor at very low energies. 
Also, we are in the process of measuring $\epsilon_{eh}$ and $F$ at different temperatures to experimentally observe the temperature dependence of these quantities.


\begin{acknowledgments}
This work was supported by Fermilab under DOE Contract No.\ DE-AC02-07CH11359. 
This manuscript has been authored by Fermi Research Alliance, LLC under Contract No. DE-AC02-07CH11359 with the U.S.~Department of Energy, Office of Science, Office of High Energy Physics. 
The CCD development work was supported in part by the Director, Office of Science, of the U.S. Department of Energy under Contract No. DE-AC02-05CH11231. 
The United States Government retains and the publisher, by accepting the article for publication, acknowledges that the United States Government retains a non-exclusive, paid-up, irrevocable, world-wide license to publish or reproduce the published form of this manuscript, or allow others to do so, for United States Government purposes.
SU was supported in part by the Zuckerman STEM Leadership Program and DR by the National Council for Scientific and Technical Research (CONICET).
\end{acknowledgments}

\vspace{5mm}


\bibliography{aipsamp}

\end{document}